\documentstyle[pra,aps,manuscript]{revtex}
\tightenlines
\RequirePackage{epsfig}
\begin{document}
\draft
\title{Optical dipole traps and atomic waveguides based on\\
       Bessel light beams}
\author{
        Jochen Arlt and Kishan Dholakia\\
        School of Physics \& Astronomy\\
        University of St. Andrews, North Haugh\\
        St. Andrews, Fife KY16 9SS
        Scotland, UK\\ \vspace{0.5cm}
        and\\
        Josh Soneson and Ewan M. Wright\\
        Optical Sciences Center and\\
        Program for Applied Mathematics\\
        University of Arizona\\
        Tucson, AZ 85721\\
        USA}

\date{\today}
\tolerance = 1000

\newcommand{\MHz}{\:\text{MHz}}
\newcommand{\kHz}{\:\text{kHz}}
\newcommand{\nm}{\:\text{nm}}
\newcommand{\um}{\:\mu\text{m}}
\newcommand{\mm}{\:\text{mm}}
\newcommand{\cm}{\:\text{cm}}
\newcommand{\mW}{\:\text{mW}}
\newcommand{\ISat}{I_{\text{Sat}}}
\newcommand{\zmax}{z_{\text{max}}}
\newcommand{\zpeak}{z_{\text{peak}}}
\newcommand{\nK}{\:\text{nK}}
\maketitle
\begin{abstract}
We theoretically investigate the use of Bessel light beams
generated using axicons for creating optical dipole traps for cold
atoms and atomic waveguiding.  Zeroth-order Bessel beams can be
used to produce highly elongated dipole traps allowing for the
study of one-dimensional trapped gases and realization of a Tonks
gas of impentrable bosons. First-order Bessel beams are shown to
be able to produce tight confined atomic waveguides over
centimeter distances.
\end{abstract}
\pacs{03.75.Fi,03.75.-b,05.30.Jp}
\newpage
\section{Introduction}
The experimental realization of Bose-Einstein condensation (BEC)
in dilute alkali vapors  \cite{AndersonEMWC95,DavisMADDKK95} has
generated substantial theoretical and experimental research
activity \cite{DalfovoGPS99}. BEC provides an important quantum
system where one can observe various phenomena such as
superfluidity \cite{RamanOVAK00}, vortices
\cite{MadisonCWD00,MatthewsAHHWC99}, spin domains
\cite{StamperKurnMCISK99} or soliton-like behaviour
\cite{ZobayPMW99,BurgerBDESSSL99,DenschlagSFCCCDHHRRSP00}. Central to
future advances with these quantum degenerate systems is their
manipulation by external potentials. The use of light fields
offers an exciting avenue in this respect as optical potentials
can be state independent, offer an array of spatial forms and can
be rapidly switched.  They offer very good prospects for novel
dipole traps and matter-wave guides.

Previous work has been reported for optical dipole traps based on
standard Gaussian light beams\cite{StamperKurnACIMSK98,Durfee99,FreseUKASGM00}.
Free-space propagating light beams such as Laguerre-Gaussian light
beams \cite{Siegman} and Bessel light beams \cite{DurninME87} are
excellent candidates for advanced all-optical manipulation of
quantum gases. Laguerre-Gaussian light beams offer a dark hollow
central region for guiding and focusing
atoms\cite{SchifferRKZSE98,ArltHD00}. Their annular form can also be
used to realise toroidal optical dipole traps for BEC
\cite{WrightAD01}, enabling studies of persistent currents. Bessel
light beams are solutions of the scalar Helmholtz equation that
are propagation invariant. This immunity to diffraction coupled
with the small size of their central region means they offer
unique characteristics in the optical domain. In this work we
discuss the use of Bessel light beams for trapping and guiding
cold atoms at or close to quantum degeneracy. Specifically we
study the Bessel light beam optical dipole trap with the aim of
generating one-dimensional quantum gases. In particular we look at
the possible realisation of a Tonks gas of impenetrable bosons
which exhibit fermionic like excitations \cite{Tonks36}.
Furthermore, we analyse in detail the waveguiding of a matter-wave
beam along a higher-order Bessel light beam. This method could
provide an important route to realising all-optical atom
interferometers and further the loading of magnetic waveguides.

%
%
\section{Bessel light beams}
In this Section we review the basics of the generation of Bessel
light beams. Ideal forms of such beams are impossible to realise
as they would have infinite extent and carry infinite power
\cite{DurninME87}. However, finite approximations to these
beams can be realised that propagate over extended distances in a
diffraction free manner.
Holographic methods offer efficient generation of Bessel beams
\cite{TurunenVF88,VasaraTF89}. Another very efficient method to
generate an approximation to a Bessel beam is by use of a
conically shaped element termed an axicon and it is this form of
Bessel beam generation we concentrate upon in this work. This
generation method is discussed in detail elsewhere
\cite{HermanW91,ArltD00} but we review it here for completeness
and clarity in notation. Our main interest is using zeroth-order
Bessel beams for optical dipole traps and first-order Bessel beams
for atomic waveguiding, and we shall concentrate on these cases.

Bessel beams are solutions of the free-space wave equation which
propagate with unchanging beam profile along the propagation axis
which we take as $z$ in cylindrical coordinates $(r,\theta,z)$.
The electric field of a monochromatic, linearly polarized ideal
Bessel beam of order $\ell$ and frequency $\omega_L$ is
\cite{DurninME87,TurunenVF88,VasaraTF89,HermanW91}
\begin{equation}
{\bf E}(r,\theta,z,t)=\frac{{\bf x}}{2} \left (E_0 J_\ell(k_r r
)e^{i(k_zz+\ell\theta-\omega_L t)} + c.c. \right ) , \label{Jell}
\end{equation}
where $E_0$ is a scale electric field value, $J_\ell$ is the
$\ell$th order Bessel function, $\ell>0$ is the azimuthal mode
number which we take as positive for simplicity in notation, and
$k_r$ and $k_z$ are the radial and longitudinal wavevectors such
that $k^2=k_r^2+k_z^2$ with $k=\omega_L/c=2\pi/\lambda_L$.  The
zeroth-order solution $J_0$ has a central maximum surrounded by
concentric rings of roughly equal power while the higher-order
solutions $J_\ell$ have zero on-axis intensity also with
concentric rings.

The method of generating a Bessel beam by use of an axicon is
based on the observation that the Fourier transform of the Bessel
beam solution (\ref{Jell}) over the transverse plane $(r,\theta)$
is an infinitely high ring in the spatial frequency domain peaked
at $K=k_r$, where the phase varies from zero to $2\pi\ell$ around
the peak: the Bessel beams may therefore be viewed as a
superposition of plane-waves with transverse wavevectors lying on
a ring of magnitude $k_r$. A finite realization of such a beam may
be produced by passing a Laguerre-Gaussian (LG) beam of order
$\ell$ through an axicon as illustrated in Fig. \ref{Fig.one}. The
axicon is a conically shaped optical element which imparts a phase
shift $\phi_{ax}(r,\theta) = k_{ax} r$ to an incident field where
$k_{ax}=k(n-1)\gamma$, $n$ being the refractive index of the
axicon material, and $\gamma$ the internal angle of the element.
By choosing $k_{ax}=k_r$ for a specific Bessel beam the axicon
imposes the ring of transverse wavevectors characteristic of the
Bessel beams on the incident beam, and these plane-waves come
together past the axicon to produce a Bessel beam. Single-ringed
LG beams (i.e. with radial mode index $p=0$), which may be
produced using holographic elements, have electric field envelopes
(plane-wave factor $\exp(i(kz-\omega_L t))$ removed) at focus of
the form \cite{Siegman}
\begin{equation}
{\cal E}_\ell(r,\theta,z=0)=\sqrt{ \frac{2P_0}{\pi w_0^2\ell !} }
\left (\frac{2r^2}{w_0^2} \right )^{\ell/2} \exp(-r^2/w_0^2)e^{i\ell\theta}  ,
\label{LG}
\end{equation}
where $P_0$ is
the power at $z=0$ incident on the axicon, $w_0$ is the Gaussian
spot size, and $\ell$ is the azimuthal mode
number. Then a LG mode of order $\ell$ incident on a axicon
with $k_{ax}=k_r$ has the appropriate azimuthal phase variation to
produce a $\ell$th order Bessel beam past the axicon. In
particular, from geometrical considerations and for an LG mode of
size $w_0$, we expect the ring of plane-waves imposed by the
axicon to overlap spatially over a longitudinal range
$\zmax\sin(\vartheta)\approx w_0$, with $\sin(\vartheta)=k_r/k$.
This gives the estimate \cite{ArltD00}
\begin{equation}
\zmax = \frac{kw_0}{k_r}  , \label{zmax}
\end{equation}
for the range over which the plane-waves overlap and produce a
finite realization of the Bessel beam profile $J_\ell(k_rr)$ past
the axicon: The larger the input spot size $w_0$ the larger the
center portion of the transverse plane over which the actual field
approaches the ideal Bessel beam.

Numerical simulations and accompanying experiments have verified
this physical picture for Bessel beam generation using an axicon
\cite{HermanW91,ArltD00}. The numerical simulations used the
Fresnel diffraction integral to propagate the input LG field
(eq. (\ref{LG})) times the phase aberration
$\exp(i\phi_{ax}(r,\theta))$ due to the axicon, to distances
beyond the axicon.  Figure \ref{Fig.two} shows a gray-scale plot
of the field intensity $I(x,y=0,z)$ as an example of this
numerical propagation for $\ell=1, w_0=0.29$ mm, $\lambda_L=780$
nm, $n=1.5$, and $\gamma=1^\circ$, giving $k_r=6.8\times 10^4$
m$^{-1}$. For this example $\zmax =3.4$ cm and the simulation
shows that this indeed estimates the scale of the Bessel beam
propagation.  As expected for a $J_1$ beam the intensity profile
has a dark fringe at the center of width of about $w_B = 1/k_r=14.7\um$.

The evolution of the field past the axicon may also be
approximated using the method of stationary phase applied to the
Fresnel integral \cite{ArltD00}, which yields the following
expression for the field intensity $I_\ell(r,z)$ for an $\ell$th
order LG input mode
\begin{equation}
I_\ell(r,z) \approx\frac{\pi 2^{\ell+1}}{\ell !} (k_r w_0) \left (
\frac{P_0}{\pi w_0^2/2} \right ) \left ( \frac{z}{\zmax} \right
)^{2\ell+1}\exp(-2z^2/\zmax^2) \,J_\ell^2(k_r r)  , \label{Bessapp}
\end{equation}
with $\zmax$ given by Eq. (\ref{zmax}).  Figure \ref{Fig.three}
shows the on-axis intensity $I(0,z)$ versus propagation distance
$z$ past the axicon obtained using both the Fresnel integral
approach (solid line) and the approximate solution (\ref{Bessapp})
(dashed line) for the same parameters as Fig. \ref{Fig.two} except
$\ell=0$, with excellent agreement (for $\ell=0$ the intensity has
a peak on-axis). In particular, by differentiating Eq.
(\ref{Bessapp}) we find that the peak intensity evaluated over the
whole transverse plane occurs at
\begin{equation}
\zpeak = \frac{\sqrt{2\ell+1}}{2}\zmax  , \label{zpeak}
\end{equation}
which gives $\zpeak=1.7$ cm, in excellent agreement with the
results in Fig. \ref{Fig.three}. The approximate expression
(\ref{Bessapp}) is applicable over the center region of the Bessel
beam if $z>k/k_r^2$ \cite{ArltD00}. Using Eq. (\ref{zmax}) this
condition can be written as $z>\zmax^2/(kw_0^2)$: Recognizing that
$kw_0^2=2z_R$ is twice the Rayleigh range $z_R$ of the input LG
beam \cite{Siegman}, and $w_B=1/k_r$ is a measure of the width of
the central lobe of the $J_0$ Bessel beam or the central dark
fringe for a $J_1$ beam, we see that if
\begin{equation}
\frac{\zmax}{z_R}=\frac{2w_B}{w_0} \ll 1  ,
\end{equation}
then Eq. (\ref{Bessapp}) should be valid over most of the
propagation range of the Bessel beam around the position of the
peak $\zpeak$.  That is, under conditions where the central spot
or dark fringe of the Bessel beam is narrow compared to the input
LG spot size, which is what we want, the stationary phase
approximation should be valid. For the above examples
$w_B/w_0=1/19.7$, and this is typical of parameters we consider.
We shall assume this condition is satisfied hereafter and use the
approximate expression (\ref{Bessapp}) in the remainder of this
paper.

Finally, we point out one further feature of Bessel beams that
highlights their utility, namely, that their radial width
$w_B=1/k_r=1/(k(n-1)\gamma)$ is determined solely by the laser
wavelength and the axicon parameters, whereas their longitudinal
extent $\zmax=kw_0/k_r=kw_0w_B$ is also dependent on the incident
LG spot size $w_0$.  This means that the longitudinal extent and
radial confinement can be varied independently.  This is in
contrast to a Gaussian beam of equal spot $w_0$ for which the
longitudinal extent of the focus is the Rayleigh range
$z_R=kw_0^2/2$.
\section{Bessel optical dipole traps}
In this Section we examine the use of zeroth-order or $J_0$ Bessel
beams for creating elongated optical dipole traps for BECs, and
assess their utility for realizing one-dimensional trapped gases
and a Tonks gas \cite{Tonks36} of impenetrable bosons
\cite{Olshanii98,PetrovSW00}. Here we concentrate on the case of
bosonic atoms but similar considerations apply to degenerate
fermionic atoms \cite{DeMarcoJ99} in Bessel beam traps.
\subsection{Gross-Pitaevskii equation}
The Gross-Pitaevskii equation (GPE) describing the macroscopic
wave function $\psi({\bf r},t)$ of an $N$-atom BEC in an optical
dipole trap can be written as \cite{LifshitzP89}
\begin{equation}
i\hbar\frac{\partial\psi}{\partial t} =
-\frac{\hbar^2}{2M}\nabla^2\psi + V(r,z)\psi + U_0 N|\psi|^2\psi ,
\label{GPeq1}
\end{equation}
where $\nabla^2=\nabla_r^2+\partial^2/\partial z^2$ is Laplacian
which is the sum of the radial and longitudinal Laplacians, $M$ is
the atomic mass, $U_0=4\pi\hbar^2a/M$ is the effective
three-dimensional short-range interaction strength, with $a$ being
the s-wave scattering length. The potential term on the
right-hand-side
\begin{equation}
V(r,z) = \frac{\hbar\Gamma^2}{8\Delta}\left (\frac{I(r,z)}
{\ISat}\right ) \label{OptPot}
\end{equation}
describes the optical dipole potential with
$\Delta=\omega_L-\omega_A$ the laser detuning from the optical
transition frequency $\omega_A$, $\Gamma$ the natural linewidth of
the optical transition, $\ISat$ is the resonant saturation
intensity, and $I({\bf r},t)=\frac{1}{2}\epsilon_0 c|E(r,z)|^2$.
For a red-detuned laser the potential is negative and the atoms
are attracted to the regions of high intensity, whereas for a
blue-detuned laser the atoms are repelled into the low field
regions.
\subsection{Optical dipole potential}
Here we investigate the properties of an optical dipole trap
formed using a red-detuned $J_0$ beam, so that the atoms are
attracted to the intense central maximum of the intensity
distribution given by Eq. (\ref{Bessapp}) with $\ell=0$
\begin{equation}
I_0(r,z) \approx 2\pi k_r w_0 \left ( \frac{P_0}{\pi w_0^2/2}
\right ) \left ( \frac{z}{\zmax} \right
)\exp(-2z^2/\zmax^2) \, J_0^2(k_r r)  . \label{J0app}
\end{equation}
We are primarily interested in tight bound optical dipole traps so
we shall approximate the optical dipole potential (\ref{OptPot})
using the parabolic approximation to the full intensity $I_0(r,z)$
around the intensity maximum at $(r=0,z=\zpeak)$:
\begin{equation}
V(r,z) - V(0,\zpeak) \approx \frac{1}{2}M\Omega_{r0}^2\left
(r^2+\lambda^2(z-\zpeak)^2 \right ) , \label{DipPot}
\end{equation}
with
\begin{equation}
\Omega_{r0}^2 =  \exp(-1/2)
\frac{\hbar\Gamma^2}{4 |\Delta|}\frac{P_0}{M \ISat}
\frac{k}{\zmax}k_r^2 , \qquad \lambda = \frac{2 \sqrt{2}}{k w_0} =
2.83 \frac{w_B}{\zmax} .\label{lambda}
\end{equation}
A red-detuned ($\Delta <0$) $J_0$ optical dipole potential
therefore provides confinement in both the radial and longitudinal
directions. Here $\Omega_{r0}$ is the radial oscillation frequency
with corresponding ground state oscillator width
$w_{r0}=\sqrt{\hbar/M\Omega_{r0}}$, and $\lambda$ is the ratio
between the longitudinal and radial trap frequencies
$\Omega_{z0}/\Omega_{r0}=\lambda$
\cite{BaymP96,PerezGarciaMH98,KivsharA99}, which also determines the
aspect ratio between the radial and longitudinal ground state
widths $w_{r0}/w_{z0}=\sqrt{\lambda}$ (in the absence of many-body
repulsion). We remark that even for tight transverse confinement
anharmonic corrections beyond the parabolic approximation for the
longitudinal ($z$) variation of the optical dipole potential
(\ref{DipPot}) are generally required.  This is evident from Fig.
\ref{Fig.three} where the on-axis intensity variation for the
$J_0$ Bessel beam is not symmetric around the peak. For this work
we shall restrict ourselves to the parabolic approximation. We
remark, however, that anharmonic corrections to the parabolic
approximation can have pronounced effects on the trap properties
such as the condensate fraction and the frequencies of the
collective excitations that can be supported by Bessel atomic
waveguides, especially for shallow traps \cite{Martikainen01}.

So far we have ignored the effects of gravity which we assume acts
in the radial direction so the longitudinal axis of the trap is
horizontal. For tight radial confinement gravity will serve simply
to displace the origin of the radial motion, and this applies
under conditions such that $M\Omega_{r0}^2 w_{r0}^2/2 \gg M g
w_{r0}$ with $g=9.81$ ms$^{-1}$. Using
$w_{r0}=\sqrt{\hbar/M\Omega_{r0}}$, this lead to the condition on
the confinement of the ground state, $w_{r0} \ll \sqrt[3]{\hbar^2/(2 M^2 g)}$. For example, for $^{85}$Rb the radial
confinement has to be less than $0.31\um$, whereas for $^{23}$Na a
confinement $w_{r0} \ll 0.74\um$ would be sufficient. We assume
that this condition is satisfied and hereafter neglect the effects of
gravity.
\subsection{One-dimensional trapped gases}
From Eq. (\ref{lambda}) the parameter $\lambda$ is seen to
determine the anisotropy of the $J_0$ optical dipole trap.  It is
well known that in the limit $\lambda\ll 1$ highly asymmetric
cigar-shaped traps are formed which are elongated along the
longitudinal direction \cite{BaymP96,PerezGarciaMH98,KivsharA99}.
Bessel beams are exceptional in this regard as they can produce
extreme asymmetries.  To illustrate this recall that $w_0$ is the
input LG spot size {\it before} the axicon, for example
$w_0=1\mm$, so that $\lambda=4.8\times 10^{-4}$ for a
$\lambda_L=1064 \nm$. We stress that one obtains such large
asymmetries using Bessel beams without sacrificing the radial
confinement, as the radial ($w_B=1/k_r$) and longitudinal
($\zmax$) extents of the Bessel beam are independently variable.
In contrast, for a Gaussian optical dipole trap changing the
focused spot size $w_0$ also changes the longitudinal extension of
the dipole trap set by the Rayleigh range $z_R=kw_0^2/2$
\cite{Siegman}.

The limit $\lambda\ll 1$ corresponds to the regime of
one-dimensional trapped gases
\cite{PetrovSW00,PerezGarciaMH98,KivsharA99} in which the radial
variation of the macroscopic wave function is effectively frozen
as the normalized ground state mode $u_g(r)$ of the radial optical
dipole potential
\begin{equation}
E_gu_g= -\frac{\hbar^2}{2M}\nabla_r^2u_g +
\frac{1}{2}M\Omega_{r0}^2r^2u_g . \label{GPeq2}
\end{equation}
Writing the macroscopic wave function in a form reflecting the
single-radial mode nature of the solution
\begin{equation}
\psi(r,z,t)=u_g(r)\phi(z,t)e^{-iE_gt/\hbar} , \label{ground}
\end{equation}
and combining Eqs. (\ref{GPeq1}), (\ref{DipPot}), (\ref{GPeq2}),
and (\ref{ground}) we obtain the one-dimensional Gross-Pitaevskii
equation for the reduced system \cite{PerezGarciaMH98,KivsharA99}
\begin{equation}
i\hbar\frac{\partial\phi}{\partial t} =
-\frac{\hbar^2}{2M}\frac{\partial^2\phi}{\partial z^2} +
\frac{1}{2}M\lambda^2\Omega_{r0}^2(z-\zpeak)^2\phi + g
N|\phi|^2\phi , \label{GPeq1d}
\end{equation}
where
\begin{eqnarray}
g&=&U_0\int 2\pi rdr|u_g(r)|^4 \nonumber \\ &=& \frac{2\hbar^2
a}{M w_{r0}^2} = 2\hbar\Omega_{r0}a, \label{g}
\end{eqnarray}
is the effective one-dimensional short-range interaction strength
\cite{PetrovSW00}. In steady-state we set
$\phi(z,t)=\chi(z)\exp(-i\mu t/\hbar)$ in Eq. (\ref{GPeq1d}) with
$\mu$ the chemical potential of the one-dimensional system.  Then
using the Thomas-Fermi approximation \cite{PetrovSW00} for high
density in which the kinetic energy term is neglected
\cite{BaymP96}, Eq. (\ref{GPeq1d}) yields for the one-dimensional
density
\begin{equation}
\rho(z) = |\chi(z)|^2 = \frac{\mu}{g} \left (1-\frac{(z-\zpeak)^2}{z_m^2}
\right )  , \qquad |z-\zpeak|\le z_m  , \label{density}
\end{equation}
with
\begin{equation}
\mu = \left (\frac{3gN}{4} \right )^{2/3}  \left
(\frac{M\Omega_{r0}^2\lambda^2}{2} \right )^{1/3}  , \qquad z_m =
\left (\frac{3Nw_{r0}^2a}{\lambda^2} \right )^{1/3} .
\end{equation}
This density solution has a peak one-dimensional density
$\rho_{\text{peak}}=1.5(N/L_z)\propto N^{2/3}$ with a longitudinal length
$L_z=2z_m$.

The Gross-Pitaevskii equation (\ref{GPeq1d}) has previously been
investigated as a model for a one-dimensional Bose-Einstein
condensate in a number of situations, including the ground state
\cite{CarrCR00a} and dynamics \cite{VillainOL01} of cigar
shaped traps \cite{AndrewsMVDKW96,KetterleDS99,BongsBDHAES00}, dark solitons
\cite{KivsharA99,BurgerBDESSSL99,FederPCSC00,CarrBBS00}, bright solitons for
negative scattering lengths
\cite{PerezGarciaMH98,KivsharA99,CarrCR00b}, gap solitons in optical
lattices \cite{ZobayPMW99}, atom waveguides
\cite{ThywissenOZDJWP99,KeyHRSHRK00,DekkerLLTSDWP00}, and as a model Luttinger
liquid \cite{MonienLE98}. Here our goal is to highlight the
utility of $J_0$ optical dipole traps for experimental studies of
one-dimensional trapped gases. To illustrate the basic scales
involved in these elongated traps we consider the case of the
$\lambda_A = 780 \nm$ transition of Rb with $\ISat = 16$W/m$^2$
and $\Gamma = 2\pi \times 6.1 \MHz$. A far red-detuned laser at a
wavelength of $\lambda_L = 1064\nm$ is used to generate a Bessel
beam with a longitudinal extent $\zmax = 10 \cm$ and a central
spot of full-width radius $3 \um$ (corresponding to $w_B = 1.25
\um$ and consequently $w_0 = 13.6 \mm$). If a laser power of $P_0
= 5$W is used this results in a trap potential of about $-49 \mu$K
depth with radial trap frequency $\Omega_{r0} = 2\pi \times 8.8
\kHz$, radial confinement of $w_{r0}= 82 \nm$, and an aspect ratio
of $\lambda = 3.5 \times 10^{-4}$. Then for $N=10^4$ atoms and a
scattering length $a = 5 \nm$, for example, we find $L_z= 2 z_m =
1.9 \mm$, and a peak density $\rho_{\text{peak}}=6.4\times 10^4$
cm$^{-1}$. For comparison, a focused Gaussian with the same radial
confinement (and trap depth) gives a trap with an aspect ratio of
only $8 \times 10^{-2}$. We remark, however, that the power
required to obtain the same radial confinement for the Gaussian
beam is much less, typically on the order of milliwatts
\cite{StamperKurnACIMSK98}. This arises because for the Bessel beam the
trapped atoms only experience the intensity in the central peak of
the beam, and the concentric rings surrounding the peak, though
key in realizing the long extent of the Bessel optical dipole
trap, do not directly affect the atomic confinement.

Bessel beam traps will suffer from the same loss mechanisms as
other shallow far-off resonance dipole traps. Apart from heating due to
diffractive collisions with background gas \cite{BaliOGGT99}, the trap lifetime will be limited by laser-noise-induced
heating \cite{SavardOT97}. The expected loss rates due to intensity fluctuations and
pointing instability of the laser beam for a Bessel beam trap should 
be comparable to those in Gaussian beam traps with identical radial
confinement and trap depth. 


Finally, we note that the $J_0$ optical dipole trap provides a
means to accelerate an atomic trapped one-dimensional gas.
Consider a one-dimensional gas in the ground state of the $J_0$
trap peaked around $z=\zpeak=kw_0/2k_r$.  If we now start to
slowly increase the LG spot size $w_0(t)$, then the longitudinal
position of the peak density of the trapped gas should vary in
time according to $\zpeak(t)=kw_0(t)/2k_r$. Such a scheme provides
a means to impart a longitudinal velocity to an initially
stationary trapped gas.
\subsection{Realization of a Tonks gas}
In the above discussion we tacitly assumed that the system of cold
atoms formed a BEC.  However, Petrov {\it et al.}
\cite{PetrovSW00} have theoretically studied the diagram of state
for a one-dimensional gas of trapped bosons, assuming $\lambda\ll
1$, and found that a true BEC, or at least a quasi-condensate,
with concomitant macroscopic occupation of a single state, is only
attained for high enough particle numbers $N>N_*$ with
\begin{equation}
N_* = \left(\frac{M g w_{z0}}{\hbar^2}\right)^2 = \left(2
\left(\frac{a}{w_{r0}}\right)
\left(\frac{w_{z0}}{w_{r0}}\right)\right)^2 . \label{Nstar}
\end{equation}
For $N<N_*$ and temperatures $T < N\hbar\Omega_{z0}$, one obtains
a Tonks gas of impenetrable bosons for which hard core repulsion
between the bosonic atoms prevents them from penetrating through
each other in the one-dimensional system, and the system
acquires properties reminiscent of a one-dimensional system of
fermionic atoms.  This remarkable property of the Tonks gas is
related to the breakdown of the spin-statistics theorem in
one-dimension, and is reflected in the Fermi-Bose mapping for this
system first elucidated by Girardeau \cite{Girardeau60}, and applied to
atomic waveguides by Olshanii \cite{Olshanii98}. Recent theoretical
investigations of Tonks gases have shown that they can support
dark soliton structures \cite{GirardeauW00b,KolomeiskyNSQ00}, and also
that their coherence properties are significantly different from
the corresponding BEC \cite{GirardeauW00a,GirardeauWT01}. Furthermore,
recent experimental developments suggest that Tonks gases should
be realizable in magnetic atom waveguides
\cite{ThywissenOZDJWP99,KeyHRSHRK00,DekkerLLTSDWP00}, and Bongs {\it et al.}
\cite{BongsBDHAES00} have proposed a hybrid trap composed of optical
dipole trap formed with a first-order LG beam combined with
magnetic longitudinal trapping. Here we examine the utilty of
$J_0$ optical dipole traps for realizing a Tonks gas.

The highly elongated Bessel beam discussed in the previous section
($\lambda_L = 1064\nm$, $P_0 = 5$W, $\zmax = 10\cm$, $w_B=1.25\um$)
would be an ideal candidate for the experimental realization of a
Tonks gas. The low aspect ratio $\lambda = 3.5 \times 10^{-4}$ and
tight radial confinement $w_{r0} = 82\nm$ result in a high upper
boundary $N_*$ for the particle number of the Tonks gas. For the
commonly used $^{87}$Rb isotope with a scattering length $a=5$ nm
one finds $N_* = 420$. Although this is still a fairly low value
it should be possible to experimentally realize a small $^{87}$Rb
Tonks gas. However, more promising would be the use of the
$^{85}$Rb isotope, where a Feshbach resonance can be used to tune
the normally negative scattering length to positive values of
several hundred nanometers magnitude \cite{CornishCRCW00}. As
$N_*$ is proportional to the square of the scattering length even a
moderate increase to $a = 50\nm$ would make it possible to create
a larger Tonks gas, with say $N = 2000$ atoms, which should be
easily detectable.

The Bessel beam trap offers some advantages compared to alternative suggested
approaches using magnetic waveguides
\cite{ThywissenOZDJWP99,KeyHRSHRK00,DekkerLLTSDWP00} and a hybrid magnetic-optical
trap \cite{BongsBDHAES00}. Firstly, it involves only a very simple
{\it all-optical} system for which the aspect ratio of the trap may be
controlled simply by varying the Gaussian spot size incident on the
axicon. More specifically, being an all-optical, it does not
involve material surfaces as in magnetic waveguides, which can cause
matter-wave decoherence \cite{HenkelW99,HenkelPW99}. Furthermore, it
allows for the possibility of trapping multiple magnetic sublevels and the
investigation of multi-component Tonks gases, which would not be
possible in the hyrbrid magnetic-optical trap of Bongs {\it et
al.} \cite{BongsBDHAES00}.
%
\section{Bessel beam atomic waveguides}
The higher-order Bessel beams $J_{\ell}$ with $\ell>0$ have zero
on-axis intensity surrounded by intense concentric rings, and a
blue-detuned beam can trap therefore atoms radially in the dark
hollow core of the beam.  In addition, since the intensity
vanishes on-axis the higher-order beams produce negligible
longitudinal confinement (see Fig. \ref{Fig.two} for the $J_1$
Bessel beam) in comparison to the $J_0$ Bessel beam. The
higher-order Bessel beams therefore present the opportunity to
realize atomic waveguides with tight radial confinement over
distances comparable with the propagation range $\zmax$ of the
Bessel beams. Furthermore, the $J_1$ beam is special in that it
provides a parabolic radial trapping potential to lowest order
(see below), whereas the higher-order trapping potential vary as
$r^{2\ell}$ near the axis. We shall therefore concentrate on the
$J_1$ atomic waveguide.
\subsection{Atomic waveguide potential}
Here we investigate the properties of an atomic waveguide formed
using a blue-detuned ($\Delta>0$) $J_1$ beam, so that the atoms
are repelled into the zero intensity central minimum of the
intensity distribution given by Eq. (\ref{Bessapp}) with $\ell=1$
\begin{equation}
I_1(r,z) \approx 4\pi k_r w_0 \left ( \frac{P_0}{\pi w_0^2/2}
\right ) \left ( \frac{z}{\zmax} \right
)^3e^{-2z^2/\zmax^2}J_1^2(k_r r)  . \label{J1app}
\end{equation}
Substituting this expression in the optical dipole potential
(\ref{OptPot}) we write the atomic waveguiding potential in the
form
\begin{equation}
V(r,z)=\frac{1}{2}M\Omega_{r1}^2(z)\left( 4 \frac{J_1^2(k_r
r)}{k_r^2} \right ) \approx \frac{1}{2}M\Omega_{r1}^2(z)r^2
,\label{WavPot}
\end{equation}
where in the last form of the potential we used the approximation
$J_1(x)\approx x/2$ which is applicable for tight radial
confinement. Here the $z$-dependent radial oscillation frequency
$\Omega_{r1}(z)$ is given by
\begin{equation}
\Omega_{r1}(z) =  \Omega_{r1}(\zpeak)\times \left
(\frac{z}{\zpeak} \right )^{3/2}\exp(-0.75(z^2/\zpeak^2-1))  ,
\label{Omr1}
\end{equation}
with $\zpeak=\sqrt{3}\zmax/2$ the position of the peak
intensity for the $J_1$ Bessel beam, and $\Omega_{r1}(\zpeak)$
is the peak radial oscillation frequency given by
\begin{equation}
\Omega_{r1}^2(\zpeak) = \left (\frac{\sqrt{3}}{2} \right )^3
\exp(-3/2) \frac{\hbar\Gamma^2}{4 |\Delta|}\frac{P_0}{M \ISat}
\frac{k}{\zmax}k_r^2  . \label{Omr1peak}
\end{equation}
In comparison to the radial oscillation frequency of the $J_0$
Bessel beam trap we find $\Omega_{r1}(\zpeak)\approx
0.5 \Omega_{r0}$.

The $J_1$ Bessel beam therefore defines an atomic waveguide whose
radial confinement peaks at $z=\zpeak$ and varies with the
longitudinal coordinate $z$. If we take the effective length $L$
of the waveguide to be the full width at half maximum of
$\Omega_{r1}(z)$ versus $z$ then by inspection of Eq. (\ref{Omr1})
we find $L\approx \zmax$, as may have been anticipated
physically based on the properties of the Bessel beams.
\subsection{Gross-Pitaevskii equation}
To examine the properties of a BEC propagating in a $J_1$ atom
waveguide we shall approximate the waveguide as invariant along the
$z$-axis over the length $L=\zmax$ and calculate the ground
state radial mode of this system.  In general we should solve the
propagation problem of the matter wave field through the varying
atomic waveguide potential, and how the atoms are funneled into
the waveguide, but that shall be the subject of future a paper.

To proceed we consider the GPE (\ref{GPeq1}) for a BEC of momentum
$p_z$ per atom moving along the $z$-axis in a cylindically
symmetric $J_1$ Bessel atom waveguide which is invariant along the
$z$ axis. Then writing the macroscopic wave function as
\begin{equation}
\sqrt{N}\psi(r,z,t) = \varphi(r)e^{i(p_z z - \mu t)/\hbar}  ,
\end{equation}
the GPE becomes
\begin{equation}
\left (\mu-\frac{p_z^2}{2M}\right )\varphi =
-\frac{\hbar^2}{2M}\left (\frac{d^2}{dr^2} +
\frac{1}{r}\frac{d}{dr} \right )\varphi + V_0J_1^2(k_r r)\varphi +
U_0|\varphi|^2 \varphi  , \label{GP4}
\end{equation}
where $\mu$ is the chemical potential, and
$V_0= 2 M\Omega_{r1}^2(\zpeak)/k_r^2$ characterizes the strength
of the $J_1$ waveguide. The wave function over the atomic
waveguide effective length $L=\zmax$ is normalized to the number
of particles
\begin{equation}
\zmax\int_0^\infty 2\pi rdr |\varphi(r)|^2 = N  . \label{Norm1}
\end{equation}
To facilitate numerical solution of the GPE we introduce scaled
variables as follows \cite{Adhikari00}
\begin{equation}
\zeta = k_r r , \qquad \varphi(r) = \sqrt{ \frac{E_r}{|U_0|}
}\varphi(\zeta) , \label{scale}
\end{equation}
in terms of which the GPE for $\varphi(\zeta)$ becomes
\begin{equation}
\alpha\varphi = -\left (\frac{d^2}{d\zeta^2} +
\frac{1}{\zeta}\frac{d}{d\zeta} \right )\varphi + \beta
J_1^2(\zeta)\varphi + c|\varphi|^2 \varphi  . \label{GP5}
\end{equation}
Here $E_r=\hbar^2k_r^2/2M$ sets the energy scale,
$\alpha=(\mu-p_z^2/2M)/E_r$ is the scaled energy eigenvalue,
$\beta=V_0/E_r$, and $c=\pm 1$ sets the sign of the many-body
interactions. The wave function $\varphi(\zeta)$ is now normalized
as
\begin{equation}
\int_0^\infty \zeta d\zeta |\varphi(\zeta)|^2 = \eta N = n ,
\label{Norm2}
\end{equation}
where the dimensionless variable $\eta=(k_r^2|U_0|/L)/(2\pi E_r) = 4 |a|/L$.

Strictly speaking the $J_1$ atomic waveguide potential in the GPE
(\ref{GP5}) does not have bound state solutions, since any initial
wave function localized in the central minimum of the $J_1$
potential will ultimately tunnel out over the finite potential
barrier due to the first peak of the Bessel beam. However, for
tight confined atoms the tunneling time can be made arbitrarily
long, and here we ignore tunneling to lowest order. Figures
\ref{Fig.four} and \ref{Fig.five} show the results of the
numerical solution of the GPE (\ref{GP5}) for $\beta=10^3$ and
$c=1$, that is repulsive many-body interactions. These numbers
would for example describe a guide for rubidium with a radius of
$5 \um$ (corresponding to $w_B = 2.78\um$) and length $L= 5\cm$
using $40\mW$ of light at $\lambda_L = 1064 \nm$. The numerical
method used to solve eq. (\ref{GP5}) was the same as in Ref.
\cite{Adhikari00}. Figure \ref{Fig.four} shows the variation of the
scaled energy eigenvalue $\alpha$ versus the scaled number of
particles $n=\eta N$, and we see that as $n$ increases so does the
energy due to the repulsive many-body interactions.  For
$n\rightarrow 0$ the scaled energy reduces to that of the atomic
waveguide potential (see below). Figure \ref{Fig.five} shows the
scaled wave function $\varphi(\zeta)$ versus scaled radial
coordinate $\zeta$ for $n=0.88$ $(\alpha=50)$ (solid line),
$n=13.4$ $(\alpha=150)$ (dotted line), and $n=43$ $(\alpha=250)$
(dashed line). Here we see that as the scaled number of atoms
increases so does the width of the wave function, as expected
physically. Clearly, there is a limit to the allowable width of
the wave function, and hence the number of atoms, as the tunneling
out of the atomic waveguide alluded to above will become more
relevant as the wave function width approaches that of the central
minimum of the $J_1$ Bessel beam.

%
\subsection{Thomas-Fermi approximation}
In this section we discuss approximate solutions to the GPE
(\ref{GP5}) to provide a framework for the numerical solutions.
First for tight confinement, so that tunneling out of the trap is
negligible, we approximate $J_1(\zeta)\approx \zeta/2$, giving
\begin{equation}
\alpha\varphi\approx -\left (\frac{d^2}{d\zeta^2} +
\frac{1}{\zeta}\frac{d}{d\zeta} \right )\varphi + \frac{\beta
\zeta^2}{4}\varphi + |\varphi|^2 \varphi  . \label{GP3}
\end{equation}
For this to be valid the macroscopic wavefunction should not
extend beyond the first peak of the $J_1$ Bessel function at $
\zeta_c=1.8$.  Then in the limit of a small number of atoms $\eta
N\rightarrow 0$ the GPE has the Gaussian ground state solution
\begin{equation}
\varphi(\zeta)\propto \exp(-\zeta^2/\zeta_0^2) , \qquad \alpha =
\sqrt{\beta} ,
\end{equation}
with $\zeta_0=2\beta^{-1/4}$: this is only valid if
$\zeta_0<\zeta_c$, or $\beta \gg 1$.  For $\beta=10^3$ this gives
$\alpha=31.6$ in agreement with Fig. \ref{Fig.four} as $\eta
N\rightarrow 0$. In the limit of large $\eta N$ we can use the
Thomas-Fermi solution \cite{BaymP96} in which the kinetic energy
term is neglected. This yields the solution
\begin{equation}
|\varphi(\zeta)|^2 = \sqrt{\eta N\beta}\left (1-\zeta^2/\zeta_m^2
\right ) , \qquad \zeta< \zeta_m = 2\left ( \frac{\eta N}{\beta}
\right )^{1/4} , \label{TF}
\end{equation}
and $\alpha_{\text{TF}}=\sqrt{\eta N\beta}$.  This solution is only valid
if $\zeta_m<\zeta_c=1.8$, which for a given value of $\beta$
places an upper bound on the scaled number of atoms $n$
\begin{equation}
n=\eta N < \beta  .
\end{equation}
In Figs. \ref{Fig.four} and \ref{Fig.five} we have restricted
$n<50$, in accordance with the above upper bound, and the wave
functions in Fig. \ref{Fig.five} are all vanishingly small in the
region $\zeta>\zeta_c=1.8$.  As an example, for $n=43,
\beta=10^3$ the Thomas-Fermi solution predicts $\zeta_m=0.9$, which is
smaller than the spatial extent of the wave function (dashed
line) in Fig. \ref{Fig.five}, and $\alpha_{\text{TF}}=207$ in comparison
to $\alpha=250$ from the exact solution in Fig. \ref{Fig.four}.
This discrepancy between the exact and approximate solutions is
not surprising as we used the parabolic approximation to the
optical potential in the Thomas-Fermi solution.  However, the
Thomas-Fermi theory captures the trends of the solution.

%
\section{Summary and conclusions}
Bessel light beams have unusual properties in the optical domain.
They have an immunity to diffraction over extended distances and
offer an elongated and narrow central region. These features give
them significant advantages over standard Gaussian light beams for
the studies presented in this work. We have shown that
zeroth-order Bessel light beams generated by use of an axicon
offer an excellent method by which to generate optical dipole
traps for one dimensional quantum gases. Typically, the 
ratio $\lambda$ between longitudinal and transversal trap frequencies for such a trap realised with a Bessel beam is nearly two
orders of magnitude smaller than that that can be achieved with a
Gaussian beam. Furthermore, we have shown that the Bessel beam
offers a potential route for an experimental realisation  of a
Tonks gas of inpenetrable bosons. Additionally we have studied the
waveguiding properties of a quantum gas along a $J_1$ (hollow) Bessel
light beam. The non-diffracting nature and small central minimum
size here make it an excellent all-optical waveguide. Such
waveguiding could be used to realise velocity filtering of cold
atoms, all-optical atom interferometers and also offer a route to
load magnetic waveguides with quantum degenerate samples.
\vspace{0.5cm}

\noindent EMW was supported in part by the Office of Naval
Research Contract No. N00014-99-1-0806, and the Department of Army
Grant No. DAAD 19-00-1-0169. JS acknowledges support from an NSF
graduate student grant. KD acknowledges the support of the UK
Engineering and Physical Sciences Research Council and the
Leverhulme Trust.

%
\begin{figure}
\caption{Illuminating an axicon with a LG mode of order $\ell$
produces a Bessel beam of the same order within the shaded
region.} \label{Fig.one}
\end{figure}
\begin{figure}
\caption{Gray scale plot of the intensity $I(x,y=0,z)$ for an
input LG beam with $\ell=1$ showing Bessel beam formation over an
elongated distance. Other parameters are $w_0=290$ $\mu$m,
$\lambda_L=780$ nm, $n=1.5$, and $\gamma=1^\circ$, giving
$\zmax=3.4$ cm, and $\zpeak=2.95$ cm.} \label{Fig.two}
\end{figure}
\begin{figure}
\caption{On-axis intensity variation for an input $\ell=0$ LG beam
using both the exact Fresnel integral approach (solid line) and
the expression (\ref{Bessapp}) based on the stationary phase
approximation (dashed line).  All other parameters are the same as
Fig. \ref{Fig.two}.} \label{Fig.three}
\end{figure}
\begin{figure}
\caption{Scaled energy eigenvalue $\alpha$ versus scaled particle
number $n=\eta N$ for $\beta=10^3$. Physically, $\alpha$ is the
chemical potential minus the longitudinal kinetic energy per
particle scaled to the energy $E_r$, and $\beta=V_0/E_r$
measures the depth of the Bessel optical potential scaled to
$E_r$.} \label{Fig.four}
\end{figure}
\begin{figure}
\caption{Scaled transverse eigenmode $\varphi(\zeta)$ versus
scaled radial coordinate $\zeta$ for $\beta=10^3$ and $n=0.88$
$(\alpha=50)$ (solid line), $n=13.4$ $(\alpha=150)$ (dotted line),
and $n=43$ $(\alpha=250)$ (dashed line).} \label{Fig.five}
\end{figure}
\newpage
\includegraphics*{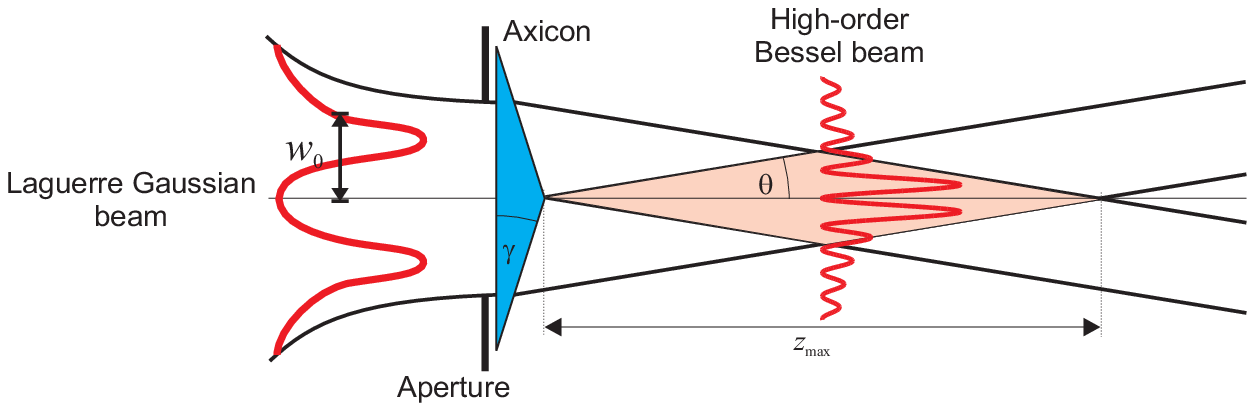}
\newpage
\includegraphics*[width=0.6\columnwidth]{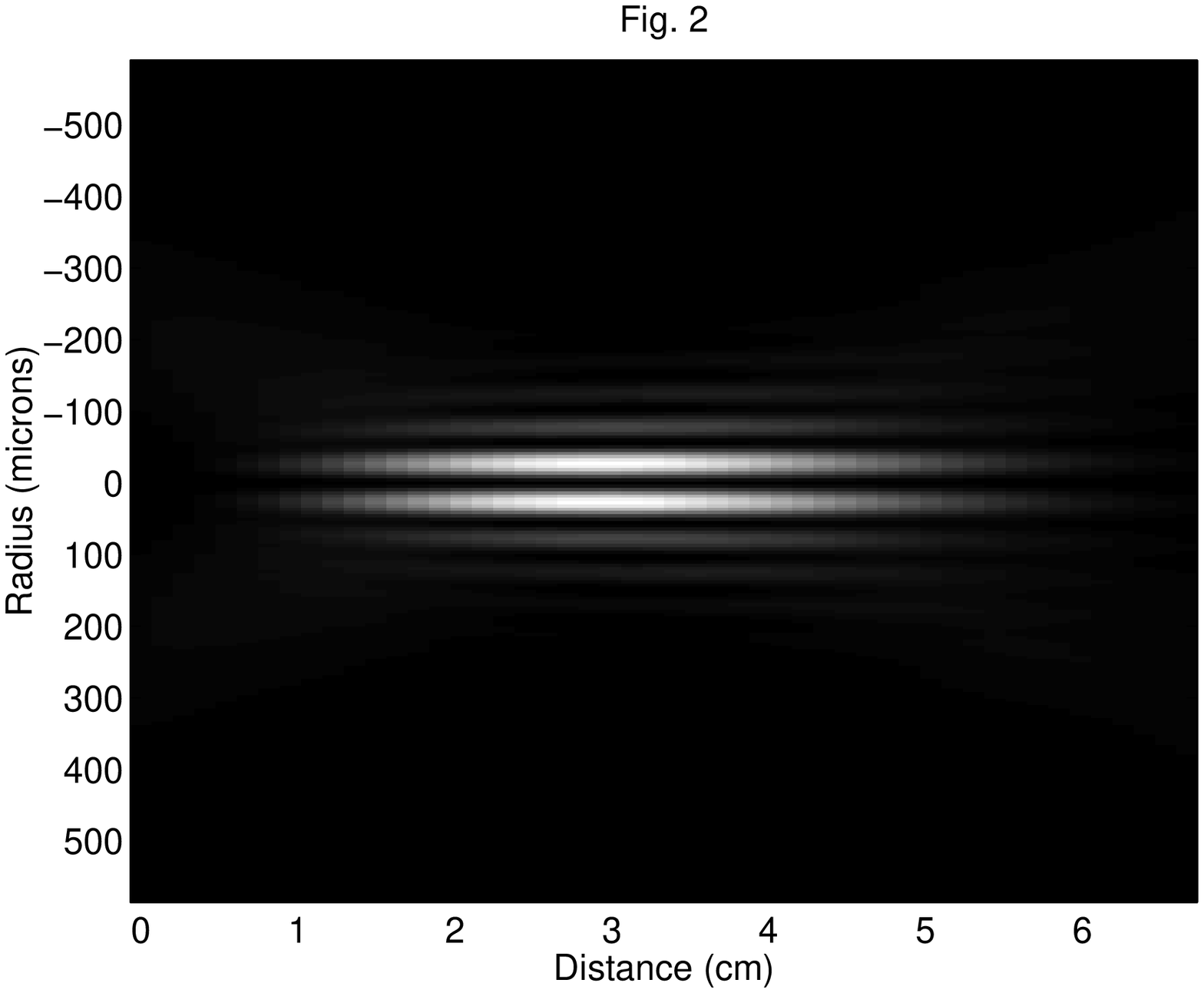}
\newpage
\includegraphics*[width=1.0\columnwidth]{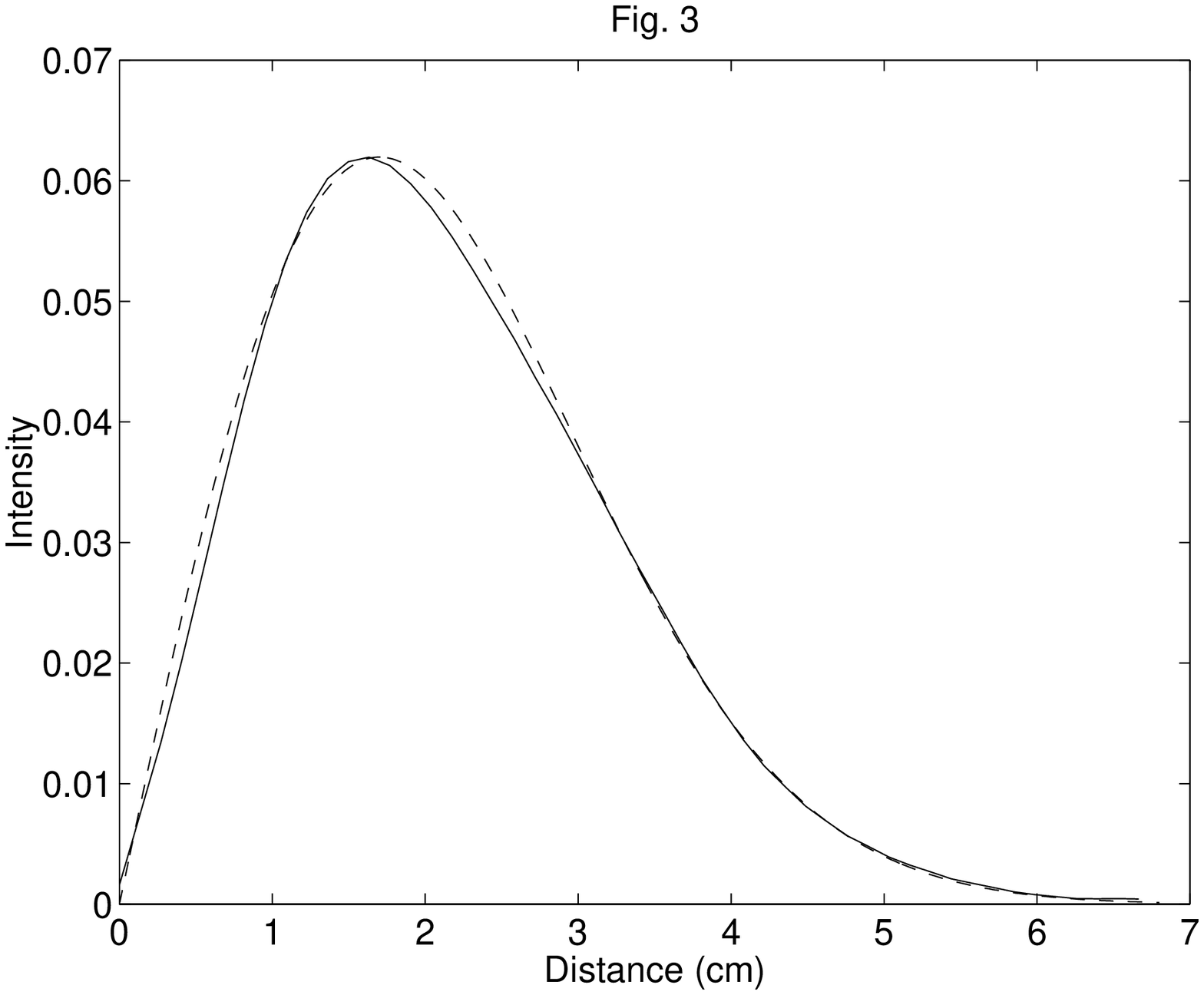}
\newpage
\includegraphics*[width=0.6\columnwidth]{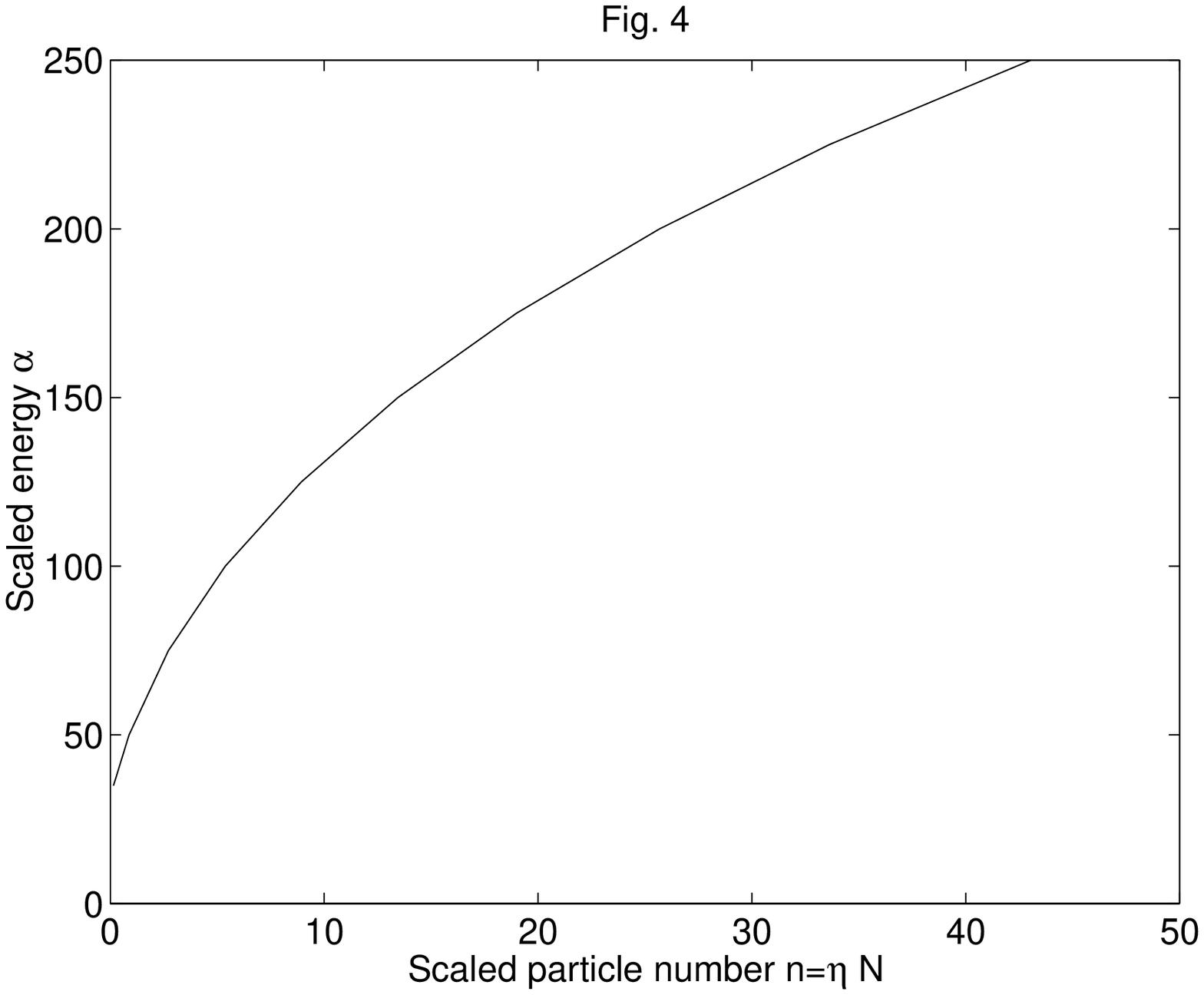}
\newpage
\includegraphics*[width=1.0\columnwidth]{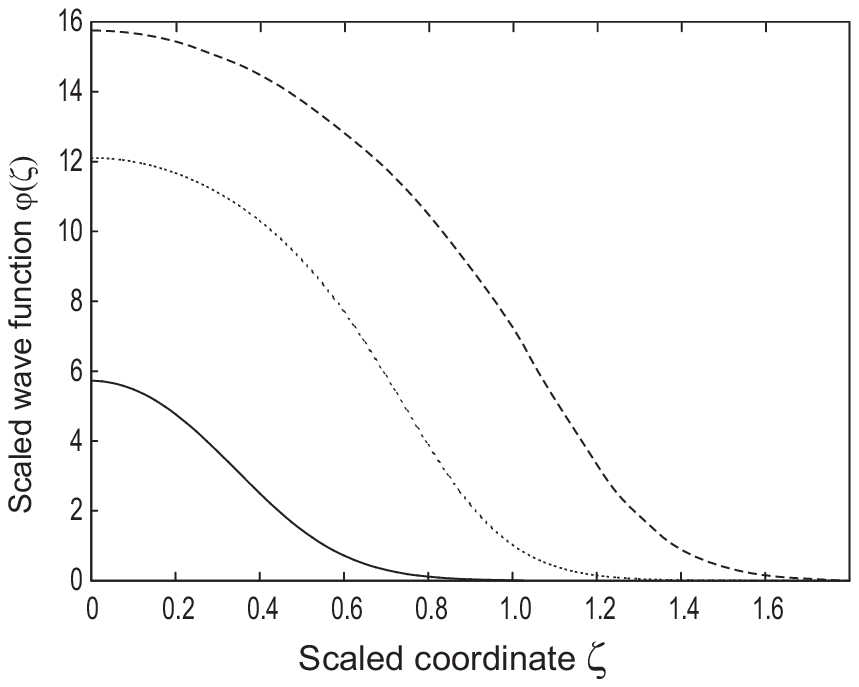}
\end{document}